\documentclass[5p]{elsarticle}
\usepackage[english]{babel}
\usepackage{amsmath}
\usepackage{siunitx}
\usepackage{graphicx}
\usepackage{lmodern}

\begin{document}
\title{High quality ultrafast transmission electron microscopy using resonant microwave cavities}
\address[tue]{Department of Applied Physics, Coherence and Quantum Technology Group, Eindhoven University of Technology, P.O.~Box 513, 5600 MB Eindhoven, The Netherlands}
\address[fei]{Thermo Fisher Scientific, Achtseweg Noord 5, 5651 GG Eindhoven, The Netherlands}
\author[tue]{W. Verhoeven\corref{cor}}
\cortext[cor]{Corresponding author}
\ead{w.verhoeven1@tue.nl}
\author[tue]{J. F. M. van Rens}
\author[fei]{E. R. Kieft}
\author[tue]{P. H. A. Mutsaers}
\author[tue]{O. J. Luiten}
\date{\today}

\begin{abstract}
Ultrashort, low-emittance electron pulses can be created at a high repetition rate by using a TM$_{110}$ deflection cavity to sweep a continuous beam across an aperture. These pulses can be used for time-resolved electron microscopy with atomic spatial and temporal resolution at relatively large average currents. In order to demonstrate this, a cavity has been inserted in a transmission electron microscope, and picosecond pulses have been created. No significant increase of either emittance or energy spread has been measured for these pulses.

At a peak current of $814\pm2$~pA, the root-mean-square transverse normalized emittance of the electron pulses is $\varepsilon_{n,x}=(2.7\pm0.1)\cdot 10^{-12}$~m~rad in the direction parallel to the streak of the cavity, and $\varepsilon_{n,y}=(2.5\pm0.1)\cdot 10^{-12}$~m~rad in the perpendicular direction for pulses with a pulse length of 1.1--1.3~ps. Under the same conditions, the emittance of the continuous beam is $\varepsilon_{n,x}=\varepsilon_{n,y}=(2.5\pm0.1)\cdot 10^{-12}$~m~rad. Furthermore, for both the pulsed and the continuous beam a full width at half maximum energy spread of $0.95\pm0.05$ eV has been measured.
\end{abstract}

\maketitle

\section{Introduction}
Ultrashort high quality electron pulses at energies ranging from 30 to 200 keV have become a useful and powerful tool to investigate dynamical systems on sub-picosecond timescales through diffraction~\cite{Sciaini2011}, imaging~\cite{Flannigan2012} or spectroscopy~\cite{VanDerVeen2015}, offering a vast amount of new information. Typically, inside an ultrafast transmission electron microscope (UTEM) electron pulses are extracted from a photocathode using an intense pulsed laser. Accurately timed with a clocking laser pulse, dynamic processes can then be investigated with pump--probe measurements. Using photoemission, a very large operational parameter-space can be spanned~\cite{Plemmons2017}. Furthermore, by using sideways illumination of a Schottky emitter, the emission characteristics of the source are maintained, allowing for high quality electron pulses to be created~\cite{Feist2017}.

Although photoemission is commonly used in UTEM systems, there is an interesting alternative to use a blanking method, where a continuous beam is periodically swept across a slit or aperture~\cite{Oldfield1976,Ura1978,Weppelman2018}. Creating pulses in this way has the advantages that amplified laser systems are no longer required, and that no intrusive alterations to the source have to be made. Instead, the system benefits from the vast amount of research done on state-of-the-art continuous sources, including recent developments that promise a higher brightness in the future~\cite{Zhang2016}. Furthermore, any possible instabilities in electron emission due to the intrinsic pointing stability of a drive laser are circumvented.

\begin{figure}
\centering
\includegraphics{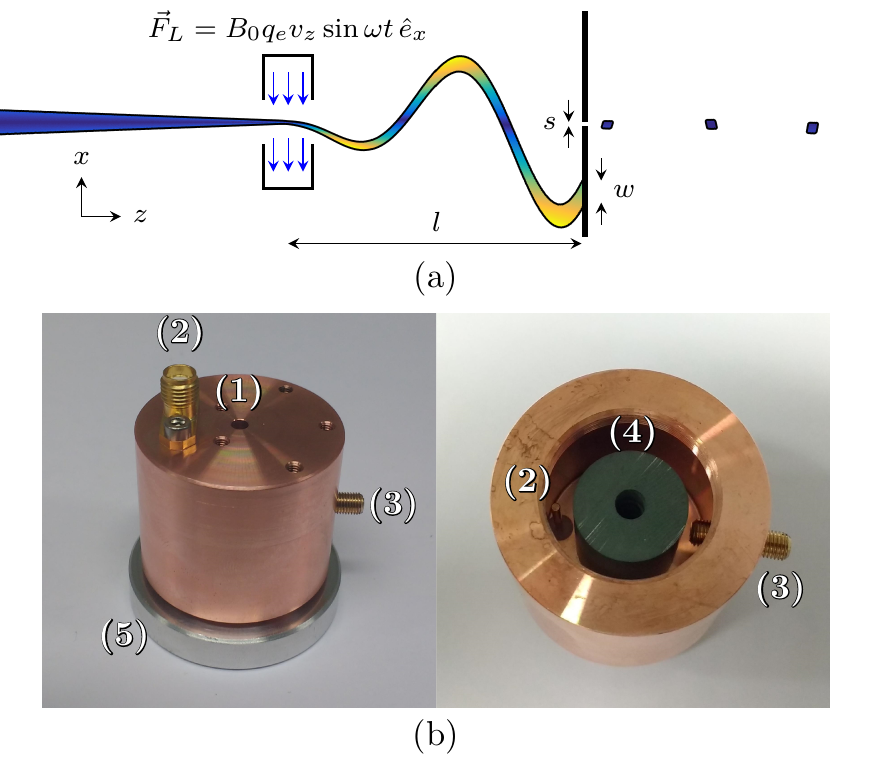}
\caption{(a) General principle of the creation of pulses using a TM$_{110}$ deflection cavity, where a continuous beam is deflected over a chopping aperture. Definition of the parameters is discussed in section \ref{section:chopping}. (b) A typical cavity, with (1) the entrance aperture, (2) the antenna, (3) the tuning stub, (4) the dielectric material, and (5) the lid used to close the cavity. Shown left is the side of the cavity, and right is the bottom of the cavity with the lid removed.}\label{fig:method}
\end{figure}

Recently, it has been shown that pulsing a beam can be done using a microwave cavity oscillating in the TM$_{110}$ mode while maintaining the low emittance of a continuous source~\cite{ThesisAdam,Verhoeven2016,VanRens2018}. This can be accomplished using a conjugate blanking scheme, where the electron beam is focused at the center of the cavity, allowing for 100~fs pulses to be created with a high beam quality. Since the power in the cavity can easily be adjusted, the pulse length can also be changed without influencing the electron emission process. In Fig.~\ref{fig:method}(a) this chopping principle is shown.

In order to perform pump--probe experiments, the phase of these microwave cavities can be accurately synchronized to a pump laser pulse. Using state-of-the-art synchronization schemes, timing jitter between the electron pulses and the laser pulses can be suppressed to levels well below 100~fs~\cite{Brussaard2013,Walbran2015}.

Alternatively, it has been proposed to use a microwave signal as a pump pulse to drive electronic or semiconductor devices for laser-free stroboscopic imaging with repetition rates in the GHz regime~\cite{Qui2015}. This is an interesting aspect of using microwave cavities, as they can provide a higher repetition rate and therefore a higher average current for samples with a fast relaxation time.

For samples with slower relaxation times, it has been proposed to use two perpendicular deflecting modes at different frequencies, which can be placed in a single cavity~\cite{ThesisAdam}. Electrons will then be created at the difference frequency of these modes, allowing for the repetition rate to be lowered to tens of MHz. If lower frequencies are desired, a fast beam blanker can be used to pick specific pulses, which are now separated by tens of ns. In this way, microwave cavities can also provide lower repetition rates for samples with slow relaxation times, allowing for the repetition rate of the setup to be optimized for each experiment.

In order to facilitate the implementation in a TEM column, these deflection cavities can be filled with a dielectric material, which allows for a reduction in both the size and power consumption~\cite{Lassise2012}. Figure~\ref{fig:method}(b) shows a typical dielectric filled cavity used for chopping an electron beam. Shown to the left is the outside of the cavity, and to the right is a bottom view of the cavity with the lid removed.

In this paper, the implementation of a TM$_{110}$ deflection cavity in a TEM is presented. Design considerations are discussed, and the performance of a cavity-based UTEM is demonstrated.

\section{Theory}
\subsection{Brightness}
An important figure of merit for a charged particle beam is its current density per unit of solid angle, called the transverse brightness. As the solid angle subtended by the beam, and therefore the brightness, depends on the beam energy, the beam quality is often expressed in terms of the \emph{reduced} brightness, which can be defined in differential form as~\cite{HawkesKasper}
\begin{equation}
B_r=\frac{1}{V^\ast}\frac{\partial^2 I}{\partial{}A \partial \Omega}\,,
\end{equation}
with $I$ the current through an area $A$ at a solid angle $\Omega$, and $V^\ast=(1/2 + \gamma/2)V$ the acceleration voltage $V$ multiplied by a relativistic correction term, with $\gamma$ the Lorentz factor. The reduced brightness is a conserved quantity during acceleration of the electrons.

Since the differential reduced brightness varies throughout the beam, its maximum on-axis value is often used, called the axial or peak brightness. Within the typical working regime of a microscope, a large portion of the emitted electrons is cut away at the condenser aperture, leading to an approximately uniform current distribution. After focusing the beam at semi-angle $\alpha$, this then results in a uniform angular distribution and a Gaussian position distribution within the beam waist, so that the peak brightness can be written as
\begin{align}
B_r&=\frac{1}{V^\ast}\frac{I}{2\pi^2\alpha^2\sigma_x\sigma_y}\nonumber\\
&=\frac{|q_e|}{m_ec^2}\frac{I}{4\pi^2\varepsilon_{n,x}\varepsilon_{n,y}}\,,\label{eq:Br}
\end{align}
with $\sigma_x$ and $\sigma_y$ the root-mean-square (RMS) size of the beam waist, $q_e$ the electron charge, $m_e$ the electron mass, $c$ the speed of light, and $\varepsilon_{n,x}$ and $\varepsilon_{n,y}$ the RMS normalized emittance in the $x$ and $y$ direction respectively, given by
\begin{align}
\varepsilon_{n,x}&=\frac{1}{m_ec}\sqrt{\langle x^2\rangle\langle p_x^2\rangle-\langle xp_x\rangle^2}\nonumber\\
&\approx \frac{\gamma v_z}{c}\sqrt{\langle x^2\rangle\langle x'^2\rangle-\langle xx'\rangle^2}\,,
\end{align}
with $v_z$ the velocity, $p_x$ the transverse momentum and $x'=v_x/v_z$ the angular distribution of the particles. In this equation, $\langle\dots\rangle$ indicates the averaging over a distribution.

\subsection{Beam chopping}\label{section:chopping}
The main advantage of using a microwave cavity is that the low emittance of the continuous beam is maintained in pulsed mode. This is only the case when using the cavity in a conjugate blanking scheme, in which all electrons deflected by the cavity originate from the same virtual image. For a regular beam blanker conjugate blanking is achieved by placing a crossover in the pivot point of the blanker.

Inside a microwave cavity the fields vary rapidly compared to the transit time of the electrons, so that it is no longer possible to distinguish a single pivot point. However, it can be shown that it is still possible to maintain the virtual image by proper placement of a crossover~\cite{ThesisAdam,VanRens2018}. For a beam chopped by an on-axis aperture, the optimal longitudinal position of this crossover is at the center of the cavity.

This is also shown in Fig.~\ref{fig:method}(a), where the beam is focused at the center of the cavity. As a result, it arrives at the chopping aperture with a certain width $w$. Sweeping this beam with a magnetic field amplitude $B_0$ and an angular frequency $\omega$ over an aperture with width $s$ results in a full width at half maximum (FWHM) pulse length of
\begin{equation}
\tau=\frac{\gamma m_e(s+w)}{4|q_e|lB_0\sin\left(f\frac{\pi}{2}\right)}\,,\label{eq:tau}
\end{equation}
where $l$ is the distance to the chopping aperture, and $f=L_\text{cav}/L_\text{max}$ the fractional length of the cavity $L_\text{cav}$ compared to the maximum useful cavity length $L_\text{max}=v_z\pi/\omega$ for which electrons feel exactly half the oscillation period. From this equation it can be seen that in order to create short pulses, the focusing angle has to be small to restrain $w$ from becoming too large.

Besides deterioration of the brightness, increase of the energy spread is also an important effect that has to be considered. Unfortunately, electrons moving through a cavity will probe the off-axis electric fields of the TM$_{110}$ mode. This will not only cause the total beam energy to change, but also the energy spread to increase. Focusing the beam at the center of the cavity minimizes this increase in energy spread, but does not completely eliminate it.

It can be shown that this additional energy spread can be decreased further by using a shorter cavity length with a higher field amplitude in order to maintain the same pulse length~\cite{VanRens2018}. The tradeoff is that more power has to be dissipated by the cavity, which brings along technical difficulties. For the results shown in this paper, a shorter cavity length is chosen at the cost of pulse length.

\section{Methods}
\begin{figure}
\includegraphics{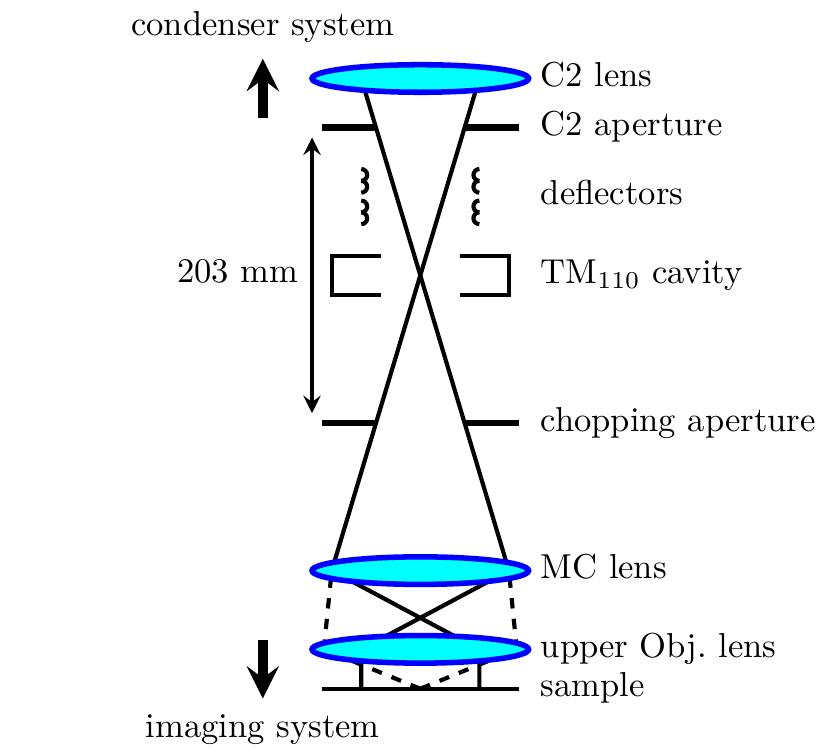}
\caption{Schematic overview of the adapted microscope column. Below the C2 aperture the column has been extended by 20.3 cm, in which a cavity, additional deflectors, and a chopping aperture have been inserted.}\label{fig:column}
\end{figure}
For the experiments, a 200 kV FEI Tecnai TEM has been elongated with a 203~mm long vacuum chamber below the C2 aperture. In this chamber a TM$_{110}$ cavity has been mounted, and an additional aperture holder has been inserted at the bottom. Above the cavity an extra set of beam deflectors has been added. Figure~\ref{fig:column} gives a schematic overview of the adapted column. The distance from the center of the cavity to the chopping aperture is $l=122.2$~mm. Both apertures are 30~\si{\micro\meter} in diameter.

In order to prevent an increase in emittance in pulsed mode, a crossover is placed at the center of the cavity by fixing the C2 lens current. The field-of-view is controlled with the minicondenser (MC) lens. This means that its original functionality of altering the divergence at the objective lens is now lost. However, the appropriate choice of apertures can mimic this functionality.

A water-cooled cavity has been designed with a resonant frequency $\omega/2\pi=2.9985$~GHz, and a length $L_\text{cav}=16.67$~mm. The cavity is loaded with ZrTiO$_4$, a dielectric material with a high permittivity and low loss tangent. The typical magnetic field amplitude in such a cavity is $B_0=1.2\pm0.1$~mT at an input power of 10~W~\cite{Verhoeven2016}. For the measurements shown in this paper, the input signal is amplified to 16~W.

As the electron beam is swept back and forth by the cavity, pulses are created twice every oscillation period. However, as these leave the chopping aperture under different angles~\cite{ThesisAdam}, half of these must be blocked. This is currently done with the SA aperture.

Using a Faraday cup, the current of the beam is measured. The energy spread of the beam is measured using a Gatan ENFINA spectrometer, with a dispersion of 0.05~eV/ch. Furthermore, measurements will be compared to particle tracking simulations using the General Particle Tracer (GPT) code~\cite{gpt}, in which realistic fields inside the cavity are taken into account, including fringe fields.

\section{Results}
\begin{figure}[t!]
\includegraphics{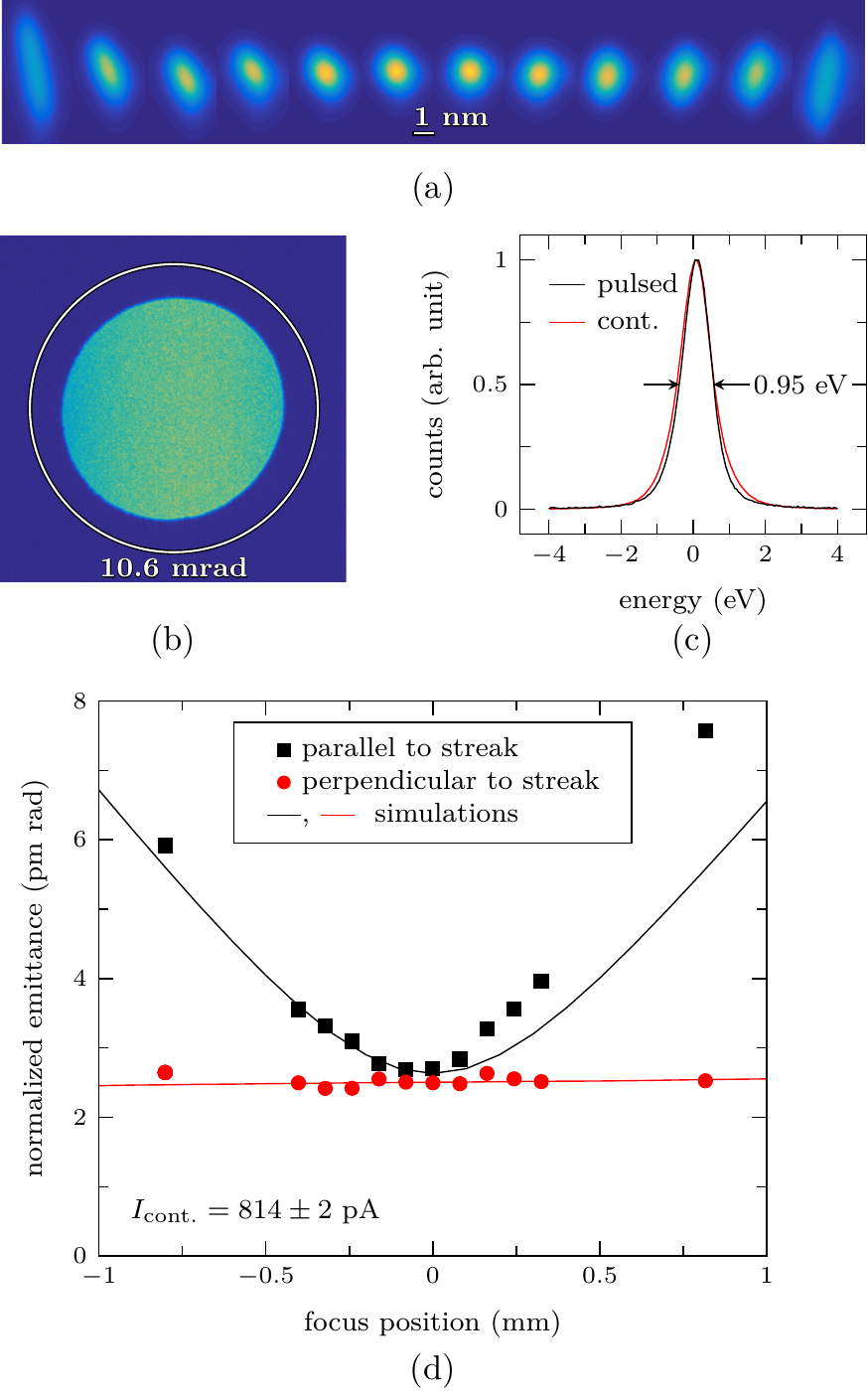}
\caption{(a) Minimal focus size of the pulsed electron beam for varying focus position within the cavity. C2 current is increased from left to right. (b)~Angular distribution of the pulsed beam, together with a known diffraction ring at 10.6 mrad used as calibration. (c)~Energy spread measured with a spectrometer for both the pulsed and the continuous beam. (d)~Emittance along the long and short axis for each spot in (a), plotted against the focus position with respect to the center of the cavity. Curves show the emittance expected from simulations.}\label{fig:emittance}
\end{figure}

Shown in Fig.~\ref{fig:emittance}(a) is the pulsed electron beam focused on the detector for varying longitudinal positions of the crossover in the cavity. Moving from left to right, the current through the C2 lens is increased, raising the focus position through the point of minimal emittance growth. At either too low or too high currents spot is elliptical, with the long and short axes corresponding to the direction parallel to and perpendicular to the sweeping direction of the cavity, respectively.

With increasing C2 current, the focused spots in Fig.~\ref{fig:emittance}(a) also rotate. This is due to the change in MC current to refocus the beam into the detector, which also rotates the beam.

Figure~\ref{fig:emittance}(b) shows the angular distribution of the pulsed beam. Angles were calibrated using a known diffraction ring at 10.6 mrad from a typical cross grating sample~\cite{AgarSample}, which is also shown in Fig.~\ref{fig:emittance}(b). From this, the focusing angle is determined to be 8.74~mrad.

In Fig.~\ref{fig:emittance}(c) the energy spread measured with the spectrometer for both the pulsed and the continuous beam is shown. The measurement with the pulsed beam seems to give a slightly lower energy spread. However, the difference is well below the resolution of the spectrometer, and is more likely to be due to small misalignments. These can easily arise between the two measurements as different settings have to be used for a continuous beam to prevent the spectrometer from saturating.

Figure~\ref{fig:emittance}(d) shows the corresponding emittance plotted against the difference in focus position. Also shown in this figure is the emittance found in simulations. These show good agreement. Deviations are attributed to the error in estimating the focus position from the lens current. The emittance of the continuous beam under the same conditions has been determined to be $\varepsilon_{n,x} = 2.5\pm0.1$~pm~rad.

From this, it can be seen that the beam quality is unaffected by the cavity in the direction perpendicular to the streak; parallel to the streak the growth of emittance can be minimized by correct placement of the crossover. At the minimum, both simulations and measurement give a negligible increase in emittance.

The minimum RMS spotsize found from these measurements is 0.61 by 0.56~nm, at a focusing angle of 8.74~mrad and a peak current of $814\pm2$~pA. From Eq.~\eqref{eq:Br}, we find a peak brightness of $6.6\cdot10^6$~A/(m$^2$ sr V). As a comparison, the RMS spotsize of the continuous beam has been measured to be 0.55 nm, resulting in a brightness of $7.5\cdot10^6$~A/(m$^2$ sr V). However, the actual brightness is presumably larger, since the measured spotsize also includes contributions from abberations.

\begin{figure}
\includegraphics{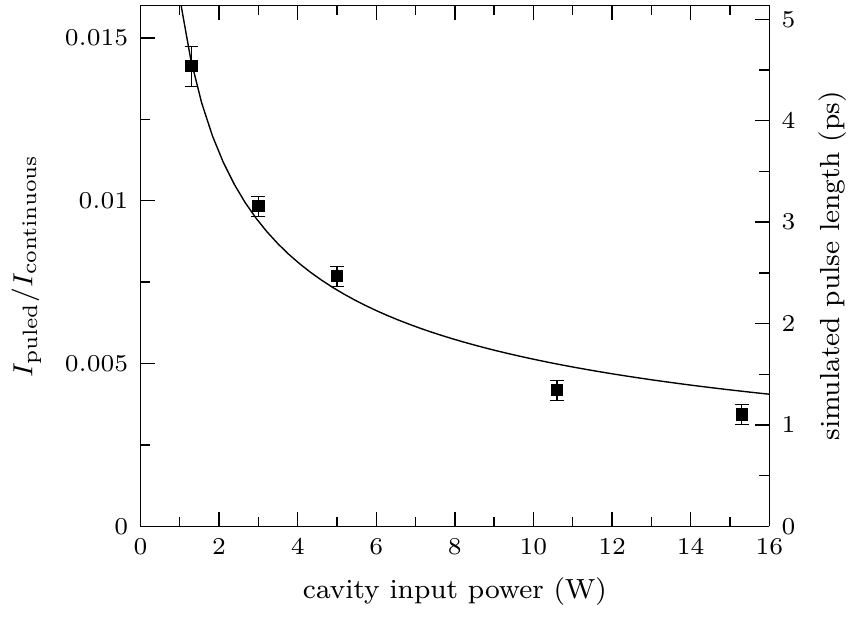}
\caption{Measured duty cycle as a function of the cavity input power, fitted with the expected behavior $\propto\sqrt{P}$.}\label{fig:current}
\end{figure}

In order to estimate the field strength in the cavity and the associated pulse length, the currents of both the continuous beam and the pulsed beam have been measured with a Faraday cup at different cavity input powers. Figure~\ref{fig:current} shows the measured current in pulsed mode divided by the continuous current on the left $y$-axis as a function of input power. The right axis shows the corresponding pulse length acquired in the simulations with the same ratio. The solid line shows a fit with the expected behavior $\propto\sqrt{P}$~\cite{Lassise2012}. At an input power of 15.3~W, a magnetic field strength of $1.45\pm0.06$~mT is expected from the fit, in good correspondence with values determined before on similar cavities~\cite{Verhoeven2016}.

At higher input powers, the measured current is smaller than expected. This could be due to measuring errors or instabilities in the electron beam, or by a change in the quality factor of the cavity as the temperature changes. In either case, a pulse length of $1.33\pm0.06$~ps is deduced from the fitted curve, whereas a pulse length of $1.1\pm0.1$~ps is deduced from the current measurement.

Shown in Table~\ref{table} are the emittance, current and energy spread measured for both the continuous and the pulsed beam. Also shown in this table are simulation results starting with a continuous beam with the same parameters, and a magnetic field strength of 1.45~mT inside the cavity. Good agreement is found between simulations and measurements.

\begin{table}
\caption{Measured parameters of the continuous beam and the pulsed beam, compared to simulations with the same continuous beam as input and a magnetic field of 1.45 mT.}\label{table}
\centering
\begin{tabular}{|l|ccc|}
\hline
&continuous&pulsed&GPT\\
\hline
$\varepsilon_{n,\parallel}$ (pm rad)&$2.5\pm0.1$&$2.7\pm0.1$&2.62\\
$\varepsilon_{n,\perp}$ (pm rad)&$2.5\pm0.1$&$2.5\pm0.1$&2.51\\
$I_\text{avg}$ (pA)&$814\pm2$&$2.8\pm0.3$&3.38\\
$\Delta E_\textsc{fwhm}$ (eV)&$0.95\pm0.05$&$0.95\pm0.05$&1.01\\
$\tau_\textsc{fwhm}$ (ps)&&&$1.31$\\
\hline
\end{tabular}
\end{table}

\section{Conclusions and outlook}
To summarize, it has been experimentally verified that TM$_{110}$ cavities can be used to create a pulsed electron beam in a 200~keV TEM without a significant increase in emittance. For pulse lengths of 1.1--1.3~ps, no measurable increase in energy spread or deterioration in performance of the microscope is found. This makes an RF-based UTEM a viable alternative to photocathodes.

As a next step, cavities will be developed further, allowing for synchronization to a clocking laser pulse at a frequency of 75~MHz. Furthermore, higher input powers will be tested, and smaller chopping apertures will be used. With increasing field strength in the cavity care must be taken to prevent an increase in energy spread. Reference~\cite{VanRens2018} explains in more detail how this can be achieved. In this way, the pulse length can be reduced towards 100~fs, allowing for pump--probe experiments with both a high temporal resolution and a high transverse coherence.

\section*{Acknowledgement}
This work is part of an Industrial Partnership Programme of the Foundation for Fundamental Research on Matter (FOM), which is part of the Netherlands Organisation for Scientific Research (NWO). The authors would like to thank E.H.~Rietman, I.~Koole, H.A.~van Doorn, and A.H.~Kemper for their invaluable technical support.

\bibliographystyle{elsarticle-num}
\bibliography{bibtex}
\end{document}